\documentclass[fleqn,usenatbib]{mnras}

\usepackage{newtxtext,newtxmath}
\usepackage{amsmath,amsfonts,amssymb}
\usepackage{graphicx}
\usepackage{setspace}
\usepackage{tocloft}
\usepackage{multicol}
\usepackage{xcolor}
\usepackage[T1]{fontenc}
\usepackage{ae,aecompl}

\usepackage{graphicx}	
\usepackage{amsmath}	
\usepackage{amssymb}	

    \title[Historic evolution of the optical design of MAORY for the ELT]{Historic evolution of the optical design of the Multi Conjugate Adaptive Optics Relay for the Extremely Large Telescope}

\author[Matteo Lombini et al.]{
Matteo Lombini,$^{1}$\thanks{E-mail: matteo.lombini@inaf.it}
Emiliano Diolaiti,$^{1}$
Mauro Patti$^{1}$
\\
$^{1}$INAF- Osservatorio di Astrofisica e Scienza dello Spazio di Bologna, via Gobetti 93/3, 40129 Bologna (Bo), Italy\\
}

\date{Accepted 2019 March 15. Received 2019 March 04; in original form 2019 February 12}

\pubyear{2019}

\begin{document}
\label{firstpage}
\pagerange{\pageref{firstpage}--\pageref{lastpage}}
\maketitle

\begin{abstract}
The optical design of the Multi Conjugate Adaptive Optics Relay for the Extremely Large Telescope experienced many modifications since Phase A conclusion in late 2009. These modifications were due to the evolution of the telescope design, the more and more accurate results of the performance simulations and the variations of the opto-mechanical interfaces with both the telescope and the client instruments. Besides, in light of the optics manufacturing assessment feed-backs, the optical design underwent to a global simplification respect to the former versions. Integration, alignment, accessibility and maintenance issues took also a crucial role in the design tuning during the last phases of its evolution. This paper intends to describe the most important steps in the evolution of the optical design, whose rationale has always been to have a feasible and robust instrument, fulfilling all the requirements and interfaces. Among the wide exploration of possible solutions, all the presented designs are compliant with the high-level scientific requirements, concerning the maximum residual wavefront error and the geometrical distortion at the exit ports. The outcome of this decennial work is the design chosen as baseline at the kick-off of the Phase B in 2016 and subsequently slightly modified, after requests and inputs from alignment and maintenance side.    
\end{abstract}

\begin{keywords}
instrumentation: adaptive optics -- telescopes -- atmospheric effects
\end{keywords}



\section{Introduction}

\label{sect:intro}  
The Multi Conjugate Adaptive Optics RelaY (MAORY)~\citep{diolaiti2008preliminary} is foreseen to be installed at the straight through focus over the Nasmyth platform of the future Extremely Large Telescope (ELT)~\citep{gilmozzi2007european}, under the coordination of the European Southern Observatory (ESO). From the optical point of view MAORY has to re-image the telescope focal plane with diffraction limited quality and low geometric distortion, for a wavelength range between 0.8 $\mu$m and 2.4 $\mu$m. From the adaptive optics (AO) point of view, MAORY must deliver a wide field correction of the atmospheric turbulence, with good and uniform Strehl ratio, accomplished with high sky coverage. Two exit ports are supposed to be fed. The first one is for the Multi adaptive optics Imaging CAmera for Deep Observations (MICADO)~\citep{davies2010micado}, placed on a gravity invariant port. MICADO requires an unvignetted field of view (FoV) of 53 arcsec x 53 arcsec, i.e a 75 arcsec circular FoV, with diffraction limited optical quality and very low geometric distortion, whose merit function is described in Section \ref{sect:dist}. The second instrument is still to be decided but anyway an exit port with a proper focal extraction and at least the same optical performance of MICADO is required. The optical interfaces at the exit ports, as the focal ratio, the exit pupil position and the focal plane curvature have been agreed with the client instruments, MICADO in particular. \\
The MAORY post focal relay (PFR) is composed by two channels, split inside the optical train: the main path optics (MPO) for the science light; the sodium laser guide stars (LGSs) \citep{foy1985feasibility} objective, that produces an image plane of LGSs used to measure the high order wavefront aberrations.\\ 
This paper is focused on the description of the MPO designs evolution,  references about the LGS objective designs can be found in~\citet{lombini2014optical,lombini2017laser}. \\
The features that are common to all MPO designs and the rationale underneath them are described in the following. 
The first feature is the requirement to deliver two clear planes in the optical path, where to place two post focal deformable mirrors (DMs). In fact, the atmospheric turbulence correction over a wide FoV implies the use of several DMs, conjugated to different atmospheric layers and described in~\citet{beckers1988increasing}. The post focal DMs number and conjugation altitude, in addition to the adaptive mirror and the field stabilization mirror (named M4 and M5) of the telescope, is a result of AO system analysis and trade-off~\citep{diolaiti2008preliminary}.
Some MPO designs with only one post-focal DM have been investigated in 2015, as a back up solution due to cost reasons.  It was anyway agreed with ESO to maintain the upgradeability to two post-focal DMs, replacing the lower conjugate DM with a rigid mirror, in case of temporary system downgrade. \\ 
The LGSs are supposed to be used for the high order modes wavefront measurements, with the scope to increase the sky coverage respect to the case of only natural guide stars (NGSs)~\citep{ellerbroek2001methods}, which are anyway necessary due to the tip-tilt and focus indetermination of the LGSs. For these low order modes (actually until the quadratic modes), the wavefront measurement is achieved by using three NGSs~\citep{rigaut2000principles}. The technical FoV is set by the sky coverage simulations and it is about 2.6-3 arcmin diameter. Since the NGS wavefront sensors (WFSs) are placed close to the exit focal ports to minimize the non-common path aberrations (NCPA) to the science instruments, all the optics size are determined by the technical FoV.\\ 
The separation of the LGS light, almost monochromatic at 589.2 nm wavelength, from the longer wavelength science light is required inside the MPO optical train. Since the LGS and NGS wavefront sensing is performed in closed loop, i.e. the WFSs are placed after the DMs, this light separation is achieved inside the optical train. The preferred solution is to use a dichroic beam splitter (dichroic hereafter), where the science light is reflected, avoiding any chromatic effect at the exit ports due to the MPO. The dichroic position is chosen to be close to a pupil plane, to allow flexibility on the LGS launching angle with very small impact on the dichroic size and to mitigate the risks on the astrometric performance caused by the high order surface irregularities (Section \ref{sect:baseline_tol}). Anyway, also different solutions, where the LGS light is separated in field (Section \ref{sect:largedm}) or the science light is transmitted through the dichroic (Section \ref{sect:backup}) have been investigated. \\
In the next sections, the most representative optical designs of the MPO produced over the years are presented. Not all designs were made available outside of the MAORY and MICADO Consortia and ESO. The designs take into account the necessary adaptation to the evolving telescope and to the interfaces variations, to both the telescope and the client instruments. When these variations were explicitly described only in confidential documents, references to ELT first-light instruments or to ELT itself papers, which consider the new requirements as a fact, are reported. The prescription data and general specifications of the old designs are indicative, since a more detailed development has been carried out only for the baseline design described in Section \ref{sect:baseline}. 
The development of the MPO to the current design considered the increasing refinement of the overall performance simulations and, as well, it was optimized to ensure feasibility of the mechanical design, the feed-backs from optics manufacturing and the more attention to the assembly and maintenance procedures. 
The acquired awareness, given by the experience, of the impact of possible variations over the global merit function used to control both performance and other requirements, permits us to be confident that the chosen baseline is a valid solution.

\section{Optical design requirements and interfaces}
\label{sect:requirements}

This section describes the main applicable requirements and interfaces relative to the optical design of the MPO. Over the years, some of them underwent to variations or refinements, driving the subsequent modifications of the designs.

\subsection{Residual wavefront error}
 
The requirement concerning the MAORY overall wavefront error (WFE) in the science FoV, which comprehends also the atmospheric turbulence residual, is about 300 nm root-mean-square (RMS) ~\citep{diolaiti2011elt}. In the WFE error budget, the allocation for the MPO optical quality is about 30-40 nm RMS, at the edge of the science FoV. Regarding the technical FoV, there is not a specified requirement, but the goal is to maintain the residual WFE below 100 nm RMS, in order to be able to calibrate it with low residuals and to avoid injecting non-atmospheric aberrations into the WF reconstruction. 

\subsection{Geometric distortion}
\label{sect:dist}
The geometric distortion requirement of the MPO derives from the astrometric requirement of MAORY~\citep{davies2010science}, whose break-down is not trivial~\citep{patti2019prep}. In the first designs, the geometric distortion was computed as the RMS of the images positions from a regular grid of sources in the scientific FoV, with the goal to be $< 0.1\%$. 
In 2014 a more detailed definition of the geometric distortion was defined with the MICADO team. It is briefly described here, more details can be found in~\citet{lombini2018optical}. Since the MPO is fixed over the Nasmyth platform, the induced geometric distortion at its exit focal ports causes the point spread function (PSF) of a point-like source, which rotates as the sky, to follow a field dependent trajectory. This field warping is caused by the non-symmetric distortion pattern due to non-symmetric off-axis mirrors. Only the Phase A design~\citep{diolaiti2008preliminary} had the symmetry to null this effect. The field derotation can compensate only the circular trajectory, while the residual part produces a PSF blur in long exposures images. In the scientific FoV, it has been agreed that the blur should not enlarge the diffraction limited PFS in the J-Band (central wavelength at 1.1 $\mu$m) of more than $1/10$ of its size. Anyway, all the other MPO designs produce low-order geometric distortions, which can be calibrated with a reduced number of sources to permit astrometric transformations of the stars coordinates. 
In late 2016 the effect of the geometric distortion in the technical FoV affecting the astrometric observations have been inserted into the merit function~\citep{patti2018maory}. The field warping is seen by the NGS WFS probes as a tilt signal, which the MCAO tries to correct. As a result, a global field shift and a plates scale variation is introduced in the scientific FoV and it produces an additional PFS blur in long exposure images~\citep{patti2019prep}. Since the absolute position of the NGS WFS probes is very difficult to be retrieved, the geometric distortion in the technical FoV can not be successfully calibrated and it has to be maintained very low by design. 

\subsection{Thermal background and throughput}
\label{sect:thermal}
The option of cooling MAORY has been discarded for cost and complexity reasons. One of the reasons is the MAORY 3 arcmin FoV. The sealing of the instrument would require a double entrance window~\citep{herriot2010nfiraos} of about 1 m diameter that would be critic from a manufacturing point of view and moreover mass issues would raise. The thermal background requirement of not exceeding 50\% of the emission of sky and telescope in the Ks band (central wavelength at  2.16 $\mu$m) depends on the assumptions of mirror reflectivity and cleanliness. On a conservative basis, if MAORY is not cooled, this requirement is reached when the MPO optical train is composed by less than 8 mirrors + dichroic. Clearly, for the efficiency of the scientific observations the minimum thermal background contribution from the MPO, and consequently the maximum throughput, is a goal.

\subsection{ELT interfaces}

The ELT primary mirror diameter was assumed to be 42 m until the beginning of 2012, when it was reduced to 39 m~\citep{mcpherson2012elt}. At the beginning of 2015 MAORY was moved from the folded focus to the straight through focus~\citep{ramsay2014elt,ramsay2016progress}. The change of MAORY position reduced the telescope reflections from six to five, but a smaller total envelope for the instrument and less distance from the telescope focal plane to the first MPO mirror were available. The optical axis of the telescope was raised from 4m to 6m above the Nasmyth platform, causing an increase of the mechanical structure weight. The instruments fixation points passed from a 1m x 1m grid to a 3m x 3m grid in middle 2016~\citep{nicklas2018micado}. This implied less flexibility for the support structure positioning over the bench for both MAORY and MICADO.\\
The ELT optical interfaces, which changed slightly during the years, are the followings: Focal ratio 17.74, exit pupil position 37.8 m toward the primary mirror, curvature at the focal plane of 9.88 m toward the primary mirror.  

\subsection{Exit ports interfaces}

\subsubsection{MICADO}

Top level requirements suggest, for the optical parameters, a 1:1 relay of the ELT focal plane delivered by MAORY. Some optical designs (Section \ref{sect:postphasea} and \ref{sect:reduced}) were compliant with this requirement, other designs did not respect it due to the trade-off with the other requirements. In the period when MICADO proposed to be directly placed at the telescope focal plane, as fall-back solution, this suggestion became a requirement. After the rejection of this option at the end of 2015, the two consortia negotiated the optical interfaces at MAORY exit ports. \\
The MICADO position changed in the different designs. When it was supposed to be attached to the MAORY bench, only the clearance for its insertion from the side was considered in the optical design. Due to stability, mass and clearance reasons, MICADO was eventually decided to be in a stand-alone configuration and its position has been conditioned by the attaching point grid of the Nasmyth platform. When in 2016 the grid spacing became coarser, less flexibility on MICADO positioning was left. Moreover, MICADO size increased over the years as described in \citet{davies2010micado,davies2016micado}.
Because of this in 2016 it was decided that its mounting and dismounting over the Nasmyth platform had to be from above and not form the side, because of its weight. Consequently, an adequate clearance was necessary to avoid to dismount any MPO optics when MICADO had to be moved. For this reason the last flat folding mirror of the MPO baseline design (Section \ref{sect:baselineopticaldesign}), which ensure the gravity invariant port to MICADO, has been chosen to be held by the MICADO structure itself. 

\subsubsection{Second instrument}

Over the years, possible candidates for the second MAORY exit port have been some spectrographs~\citep{oliva2008high,thatte2014harmoni} but, at the moment of writing, the instrument is still undefined. From the optical quality and interfaces point of view, it has been agreed that they have to be similar to the ones delivered to MICADO. Regarding the position, the possibility to have a back focal distance of more that 0.5 m has always been requested. For this reason, some optical designs have been discarded because the second port, selected by means of a deployable flat mirror, was too close to the MICADO axis (Section \ref{sect:finaltradeoffs}).     

\subsection{Internal requirements}

\subsubsection{Deformable mirrors}
\label{dm}

In Phase A the two DMs were conjugated to 12.7 km and 4 km distance from the telescope aperture while, on the basis of further simulations, the conjugation altitude of the DMs was raised to stay within a range of 15-16 km and 6-8 km respectively. DMs pitch was initially considered to be about 7-8 $\mu$m and about 50 actuators along the diameter were considered, resulting in a DMs size of about 400 mm. In order to mitigate the issues caused by the conjugation range of the tilted DMs and by the atmospheric layer image blur on the DM planes~\citep{lombini2014optical}, subsequent designs (from Section \ref{sect:largedm}) changed the actuator pitch size to 25-29 mm, on the basis of existing 8 m class telescopes secondary mirrors~\citep{crepy2010last,gallieni2010voice,riccardi2003adaptive,close2010magellan}. To maintain the same actuator number, the DMs size increased to about 1.2 m diameter but afterwards it was reduced to 700-800 mm (Section \ref{sect:reduced}), trading a little between the AO performance and overall size and cost~\citep{oberti2017maory}. DM shapes has been flat as well as curved on axis surfaces or even off-axis, as for the Giant Magellan Telescope secondary mirrors~\citep{biasi2010gmt}, with the goal of reducing the number of MPO elements (from design in Section \ref{sect:reduced}). In the baseline design (Section \ref{sect:baselineopticaldesign}), an additional effort has been done to maintain equal and spherical the curvature radii of the DM reference bodies~\citep{biasi2011contactless}, for cost reasons and calibration convenience. The different surface asphericity is planned to be achieved by the active compensation of the actuators themselves.  

\subsubsection{Mirrors}

Phase A sky coverage simulations results defined the optics size to accept an unvignetted technical FoV of 160 arcsec, where to find the NGSs. More detailed simulations suggested that 180 arcsec is a better value for the technical FoV~\citep{oberti2017maory}. The 1 m class optics that compose the MPO is almost independent by the DM diameter. Also in Phase A, where the DMs were considerably smaller (Section \ref{sect:phasea}), the optics size close to the entrance and exit focal planes of MAORY were defined by the FoV. Likewise the overall envelope of the MPO is little dependent by the DMs size. 

\subsubsection{Dichroic}
 
The dichroic size of phase A design had an elliptical shape of about 1000 mm x 650 mm. In the next designs the dichroic size was reduced to about 600 mm diameter, after some feed-backs from optical manufacturers. The dichroic is expected to have a 3\%-5\% of infrared emissivity. In the effort of reducing the MPO contribution to the overall thermal background, a design with an annular dichroic where the science light passed undisturbed has been investigated (Section \ref{sect:reduced}). Also the option where the science light passed through the dichroic, as a back-up solution, has been studied and it is briefly described in Section \ref{sect:backup}.  

\subsubsection{Other requirements}

Any realistic optical design must take into account also the coupling with the mechanical design (the main bench and the optical mounts), the assembly, integration and verification (AIV) phase, the maintenance, the sub-systems interfaces, etc. The consideration of these internal requirements increased with the convergence to the baseline design. They played a relevant role in the latest trade-off stages among different options and design modifications.

\section{Phase A design}
\label{sect:phasea}

\subsection{Design rationale}

In this design, dated 2009, the ELT diameter was 42 m, the optical axis respect to the Nasmyth platform was at 4 m height and MICADO was supposed to be mechanically held and rotated by MAORY, which was positioned after one of the folded focii of the ELT. 

\subsection{Optical design}

\begin{figure}
	\includegraphics[width=\columnwidth]{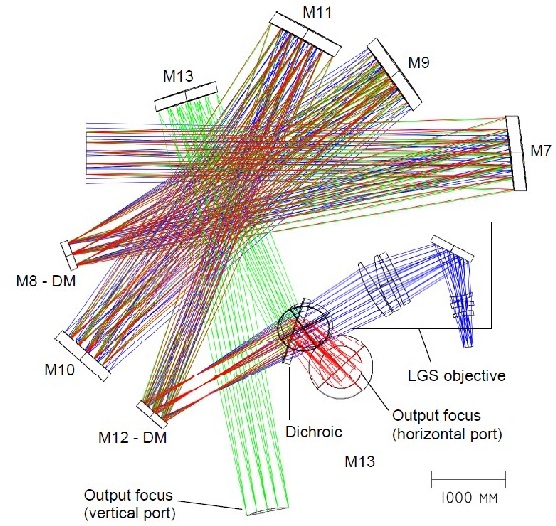}
    \caption{Phase A optical design of the MPO.}
    \label{fig:phase_A}
\end{figure}

The MPO consists of 7 mirrors plus the dichroic for MICADO and 8 mirrors plus the dichroic for the second port.
The Phase A optical design of the MPO (Figure \ref{fig:phase_A}) is based on two pairs of off-axis concave mirrors, almost parabolic. Two intermediate pupil images are produced for two flat post-focal DMs, optically conjugated to 12.7 km and 4 km range from the telescope pupil. An additional folding mirror allows to accommodate the MPO in the reserved area of the Nasmyth platform. An off-axis distance along the x-axis is introduced to raise the output focal station on the gravity invariant port. This design fulfills the WFE and geometrical distortion requirements. The average RMS blur on the DM plane is about $1/10$ of sub-aperture. Anyway, the Phase A review pointed out two major drawbacks. The first one is the considerable exit focal plane curvature radius of about 1 m due to the use of only concave mirrors; the second one concerns the dichroic size of about 1 m in one axis and its tilt respect to the normal of about 47 degrees. \\

\begin{table}
\centering
\caption{Phase A design general prescriptions and specifications.}
	\label{tab:phaseA} 
	\begin{tabular}{|l|l|l|l|}
\hline
		\textbf{Surface}   & \textbf{Diam}  &  \textbf{Curv}      & \textbf{Shape}                  \\
                  & [mm]      &                 &                         \\
		\hline
        M7        &  1030      &  Concave       &  Off-axis paraboloid    \\
        M8   &   350      &  Flat          &                         \\
        DM@12.7 km&         &                &   \\
        M9        &  1150      &  Concave       &  Off-axis paraboloid    \\
        M10       &   1000    &    Flat         &                          \\
        M11       &    1000   &  Concave        &    Off-axis paraboloid   \\
        M12  &    425    &    Flat         &                          \\
        DM@4 km&         &                &   \\
        dichroic & 950 x 650  &    Flat         &                          \\ 
        M13      &    870     &  Concave        &     Off-axis paraboloid   \\ 
        M14      &    950 x 650     &  Flat       &       \\ 
       (for 2nd port) &  &  &  \\
        \hline
\multicolumn{4}{|c|}{\textbf{Optical interfaces to exit port} }                          \\ \hline
\multicolumn{2}{|l|}{Focal ratio} & \multicolumn{2}{l|}{F/17.7} \\
\multicolumn{2}{|l|}{Exit pupil distance}   & \multicolumn{2}{l|}{37000 mm (towards telescope)} \\ 
\multicolumn{2}{|l|}{Focal plane curvature}   & \multicolumn{2}{l|}{1500 mm (convex to telescope)} \\ 
\multicolumn{2}{|l|}{NGS patrol FoV}   & \multicolumn{2}{l|}{200 arcsec diameter} \\ \hline
\end{tabular}
\end{table}

\section{Post-phase A design}
\label{sect:postphasea}

\subsection{Design rationale}

\begin{figure}
\begin{center}
\includegraphics[width=\columnwidth]{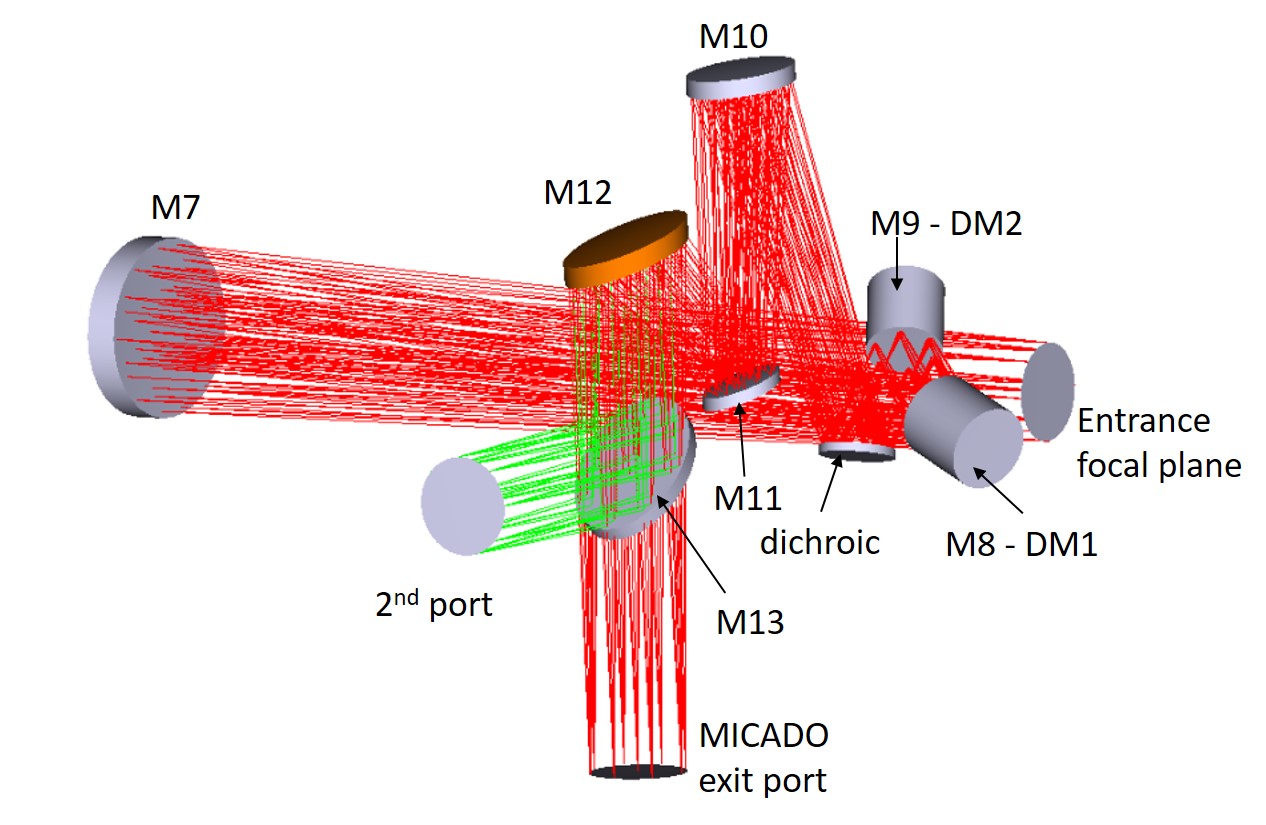}
\end{center}
\caption{Post Phase A optical design of the MPO. } 
\label{fig:post_phase_A}
\end{figure} 

In the post-phase A (2009-2011), a new optical design was developed~\citep{diolaiti2011elt} to solve the issues concerning the focal plane curvature and the dichroic size, raised at the Phase A review. 
\subsection{Optical design}
The MPO consists of 6 mirrors plus the dichroic for MICADO and 7 mirrors plus dichroic for the second port.
In this design, the first off-axis mirror creates a collimated beam, where the two post-focal DMs and after the dichroic are placed. The beam is then focused by a 3-mirrors assembly, including two concave mirrors and one convex mirror, to control the exit port field curvature. The MPO delivers the same optical interfaces of the telescope, the WFE and geometrical distortion fulfill the requirements. Anyway, for clearance reasons, the two DMs tilts respect to the optical axis causes a conjugation range between the mirror edges of $1\pm 2$ km and the optical quality of the layers to the DM planes is too low, about $1/5$ of sub-aperture size. 

\begin{table}
\centering
\caption{Post Phase A design general prescriptions and specifications.}
	\label{tab:postphaseA} 
	\begin{tabular}{|l|l|l|l|}
\hline
		\textbf{Surface}   & \textbf{Diam}  &  \textbf{Curv}      & \textbf{Shape}                  \\
                  & [mm]      &                 &                         \\
		\hline
        M7        &  1250      &  Concave       &  Off-axis ellipsoid    \\
        M8 &   500      &  Flat          &                       \\
        DM@12.7km &            &                &                       \\
        M9  &  450     &  Flat       &                              \\
        DM@4km   &           &            &                               \\
       dichroic &   460      &    Flat         &                      \\ 
        M10  &    650    &   Concave         &     Off-axis hyperboloid $\ast$ \\
        M11       &  460     &  Convex        &   Off-axis hyperboloid $\ast$ \\
        M12       &  740     &  Concave        &   Off-axis hyperboloid $\ast$ \\
        M13       &  950 x 650     &  Flat        &                     \\
       (for 2nd port)  &       &                  &                     \\
        \multicolumn{4}{|l|}{$\ast$ even aspheric terms present }                          \\ \hline
\multicolumn{4}{|c|}{\textbf{Optical interfaces to exit port} }                          \\ \hline
\multicolumn{2}{|l|}{Focal ratio} & \multicolumn{2}{l|}{F/17.7} \\ 
\multicolumn{2}{|l|}{Exit pupil distance}   & \multicolumn{2}{l|}{37000 mm (towards telescope)} \\ 
\multicolumn{2}{|l|}{Focal plane curvature}   & \multicolumn{2}{l|}{9900 mm (convex to telescope)} \\ 
\multicolumn{2}{|l|}{NGS patrol FoV}   & \multicolumn{2}{l|}{200 arcsec diameter} \\ \hline
\end{tabular}
\end{table}

\section{Designs with large DM actuators spacing}
\label{sect:largedm}
\subsection{Designs rationale}
At the beginning of 2012, when the ELT diameter was reduced to 39 m~\citep{mcpherson2012elt} and MAORY was still at the folded focus of the Nasmyth platform, the MPO design switched to include bigger DM pitches (25 mm - 28 mm), in order to mitigate the issues raised in the post Phase A design (Section \ref{sect:postphasea}) on the layers conjugation ranges of the DMs and on the optical quality at the DM planes. This pitch technology was already in operation at the Large Binocular Telescope~\citep{riccardi2010adaptive} and studies for other telescopes were being carried out~\citep{miller1999robust,biasi2010gmt}. Hereafter, all the MPO designs have the DMs with this inter-actuator distance class. \\
To maintain the number of actuators similar to the Phase A case (Section \ref{dm}), the DMs size increased to about 1.1 m, the same size of the adaptive secondary mirrors of the Very Large Telescope \citep{arsenault2006deformable}. The issued related to the optical quality on the DM planes as well the DM conjugation range were mitigated but still the dichroic size issue was present. If both DMs were flat, the dichroic size was about 1 m. For this reason the MPO optical design evolved to have and intermediate focal plane, as described in the following. \\
In the next designs, the intermediate focal plane is produced by the second DM, supposed to be an off-axis ellipse~\citep{hinz2010gmt}, or by a rigid off-axis concave mirror after the DMs. The goal was to produce a smaller pupil image, by means of a collimator, where one can insert the dichroic, reducing the size of this component. The exit ports optical parameters depended on the number of optical elements after the dichroic. This design is shown in Figure \ref{fig:intermediate}.\\ 
Field splitting can be alternatively achieved by placing a perforated dichroic on the intermediate focal plane (Figure \ref{fig:annular}), where the LGS light is reflected and the NGS light is refracted. No additional background contribution is added to the science beam, as the light passes through the central clear hole.
With this field splitting configuration, the science beam passes through the central hole in the perforated dichroic, while the NGS light, in the technical FoV, is transmitted through the annular portion of the dichroic itself. Finally, the LGS light was reflected on the annular part of the dichroic. The requirement to let the science field pass through the central hole posed limitations both on the NGS technical FoV and on the LGS launching angle. Considering a maximum science FoV of 53 arcsec x 53 arcsec, the minimum LGS launching angle must be $\ge$ 1.7 arcmin. The NGSs chief rays were subject to a shift and a tilt respect to the optical axis when passing through it. A second refractive elements was thus necessary to align the NGS FoV axis of rotation to the science one and to compensate the WFE of the NGS beams by the dichroic. \\
Among the different variants on this design class, Figures \ref{fig:intermediate} and  \ref{fig:annular} show the cases with the minimum number of optical elements to reduce the thermal background contribution (Section \ref{sect:thermal}).

 \subsection{Optical designs}

\begin{figure}
\begin{center}
\includegraphics[width=\columnwidth]{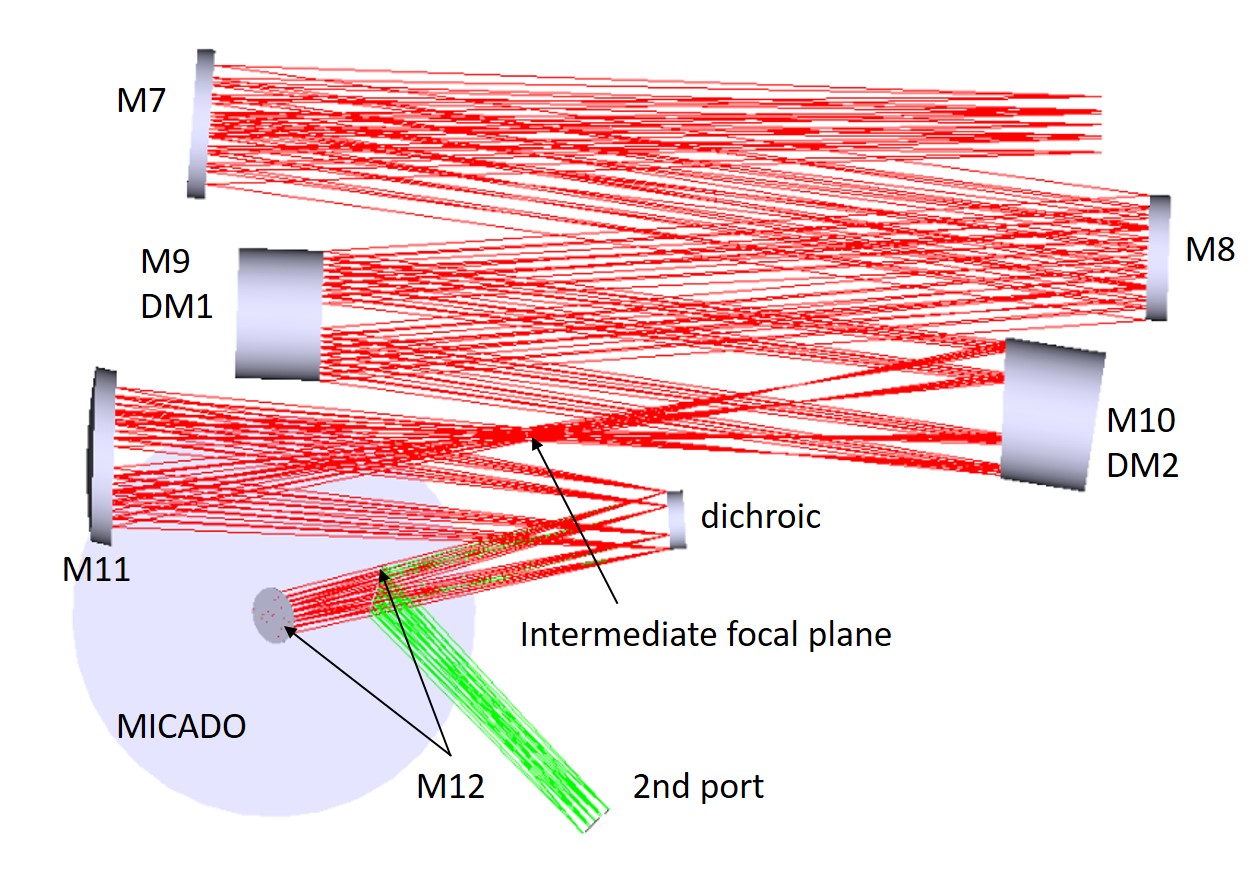}
\end{center}
\caption{Intermediate focus optical design of the MPO. In this version the dichroic is placed after the collimating mirror M11.} 
\label{fig:intermediate}
\end{figure}

\begin{figure}[!ht]
\begin{center}
\includegraphics[width=\columnwidth]{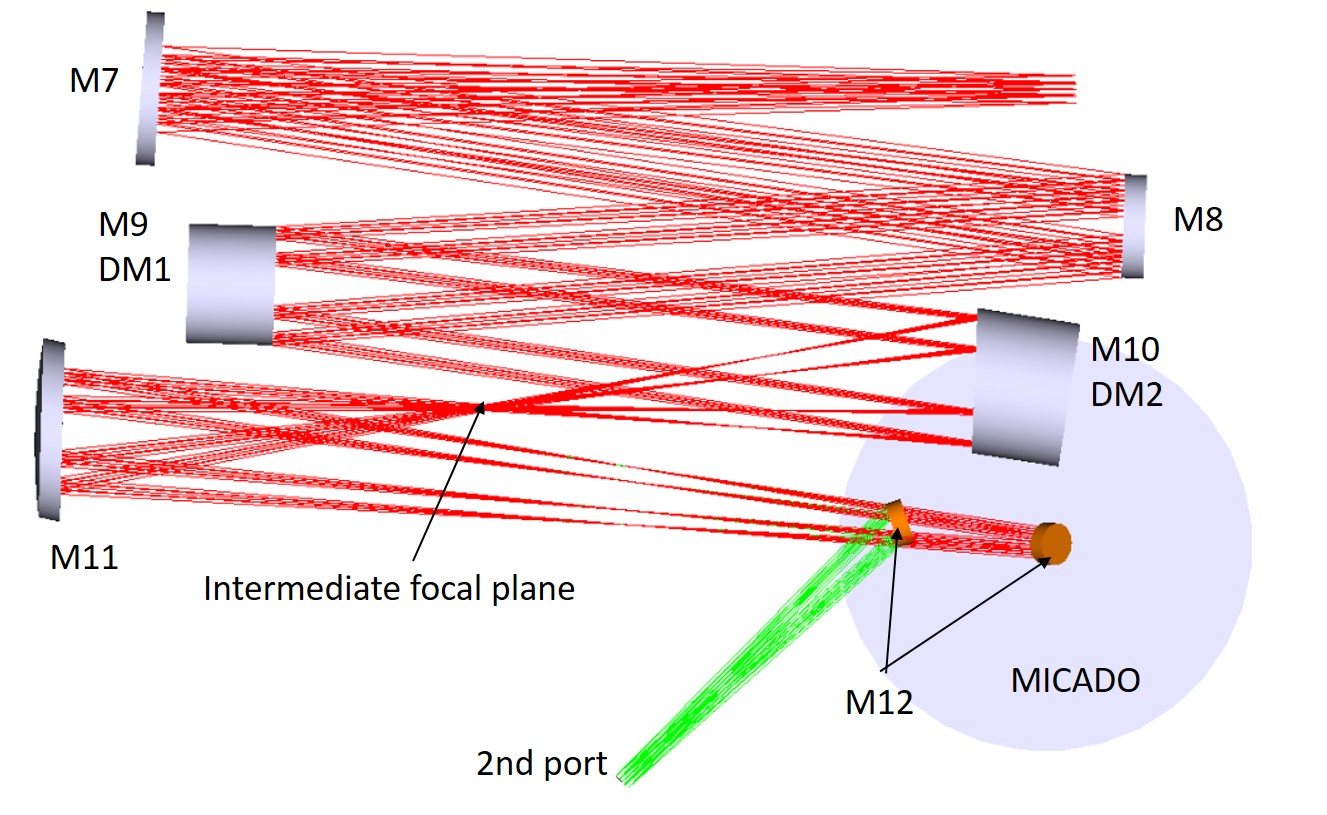}
\end{center}
\caption{Intermediate optical design of the MPO. In this version the annular dichroic is placed at the intermediate focal plane after the second DM.} 
\label{fig:annular}
\end{figure}

The MPO consists of 6 mirrors for MICADO (plus the dichroic in the case described in Figure \ref{fig:intermediate}) and one deployable flat mirror more for the second port.
The first  mirror (M7) forms a pupil image. Flat mirror M8 is necessary to fit the allowed space~\citep{lombini2014optical}. The DM1 is flat and DM2 is an off-axis ellipsoid, which produces a focal plane. In this focal plane a perforated dichroic can be placed. A concave mirror (M11) forms a pupil image. The two designs differ only for the dichroic type and position: in Figure \ref{fig:intermediate}, it is placed in the pupil image; in Figure \ref{fig:annular}, there is a perforated dichroic in the intermediate focal plane. They have the same optical interfaces at the exit ports, which anyway differ from the telescope, especially regarding the field curvature due to the use of only concave mirrors.   

\begin{table}
\centering
\caption{Intermediate focus design general specifications and  prescriptions.}
	\label{tab:intermediate} 
	\begin{tabular}{|l|l|l|l|}
\hline
		\textbf{Surface}   & \textbf{Diam}  &  \textbf{Curv}      & \textbf{Shape}                  \\
                  & [mm]      &                 &                         \\
	\hline
	     M7        &  1250      &  Concave       &  Off-axis hyperboloid    \\
        M8        &  1200      &  Flat       &                           \\
        M9   &   1200      &  Flat          &                         \\
        DM@12.7km &            &                &                       \\
        M10 &  1200      &  Concave       &  Off-axis ellipsoid    \\
        DM@4.7km &            &                &                       \\
        M11       &  1250      &  Concave        &    Off-axis ellipsoid   \\
        dichroic &   700      &    Flat         &                          \\ 
            (if present) &       &                  &                           \\
        M12  &    500 x 700    &    Flat         &                          \\
        M13       &  400 x 600     &  Flat        &                         \\
       (for 2nd port)  &       &                  &                           \\
        \hline
\multicolumn{4}{|c|}{\textbf{Optical interfaces to exit port} }                          \\ \hline
\multicolumn{2}{|l|}{Focal ratio} & \multicolumn{2}{l|}{F/11} \\
\multicolumn{2}{|l|}{Exit pupil distance}   & \multicolumn{2}{l|}{5800 mm (towards telescope)} \\ 
\multicolumn{2}{|l|}{Focal plane curvature}   & \multicolumn{2}{l|}{1900 mm (convex to telescope)} \\ 
\multicolumn{2}{|l|}{NGS patrol FoV}   & \multicolumn{2}{l|}{160 arcsec diameter} \\ \hline
\end{tabular}
\end{table}

\section{Towards the MAORY Phase B}
\label{sect:reduced}
\subsection{Designs rationale}
Since the project phase A, MICADO proposed a risk mitigation path, to be activated in case of delays of the MAORY project, in which MICADO would be operated in stand-alone mode, with a Single-Conjugate Adaptive Optics (SCAO) system consisting of a single NGS WFS and based on M4-M5 as corrective elements. At the beginning of 2015, MICADO proposed to implement this risk mitigation path, if necessary, by placing the MICADO instrument directly at the telescope focal plane. 
For this reason, it was mandatory for MAORY MPO to replicate the telescope optical interfaces. Moreover, MAORY was moved to the straight through focus of the Nasmyth platform and the optical axis was raised from 4 m to 6 m height. Besides, in view of the start of the project Phase B begin at the end of that year, a bigger attention was paid to other requirements as the integration procedure and overall cost.
The MPO plant had to be smaller than 12 m x 6 m, to fit in the integration hall and the available space on the Nasmyth platform. Regarding the DMs, it was decided to decrease their size to 700-800 mm, since the loss in performance was acceptable~\citep{diolaiti2008preliminary}, in order to save space and mass and to reduce the instrument overall cost. The DMs was designed as on-axis curved surfaces to decrease the total number of optical surfaces and maintain the dichroic size at an acceptable size. On top of that the DMs curvature was imposed to be equal to reduce manufacturing complexity and cost. The LGSs are fixed respect to the telescope primary mirror and, from the Nasmyth platform, they rotate as the telescope pupil. The post focal DMs are seen as rotating from the LGS WFSs. A possible option at that time was to mechanically rotate the post focal DMs instead of the numerical rotation of the DM interaction matrix, thus an on-axis shape were necessary. Finally, the same optical power permitted to save cost and to share the spare parts. The conic constants of the mirrors were limited to negative values, oblate ellipsoids or paraboloids or hyperboloids, to facilitate the manufacturing and testing process. 

\subsection{Optical design}

\begin{figure}
\includegraphics[width=\columnwidth]{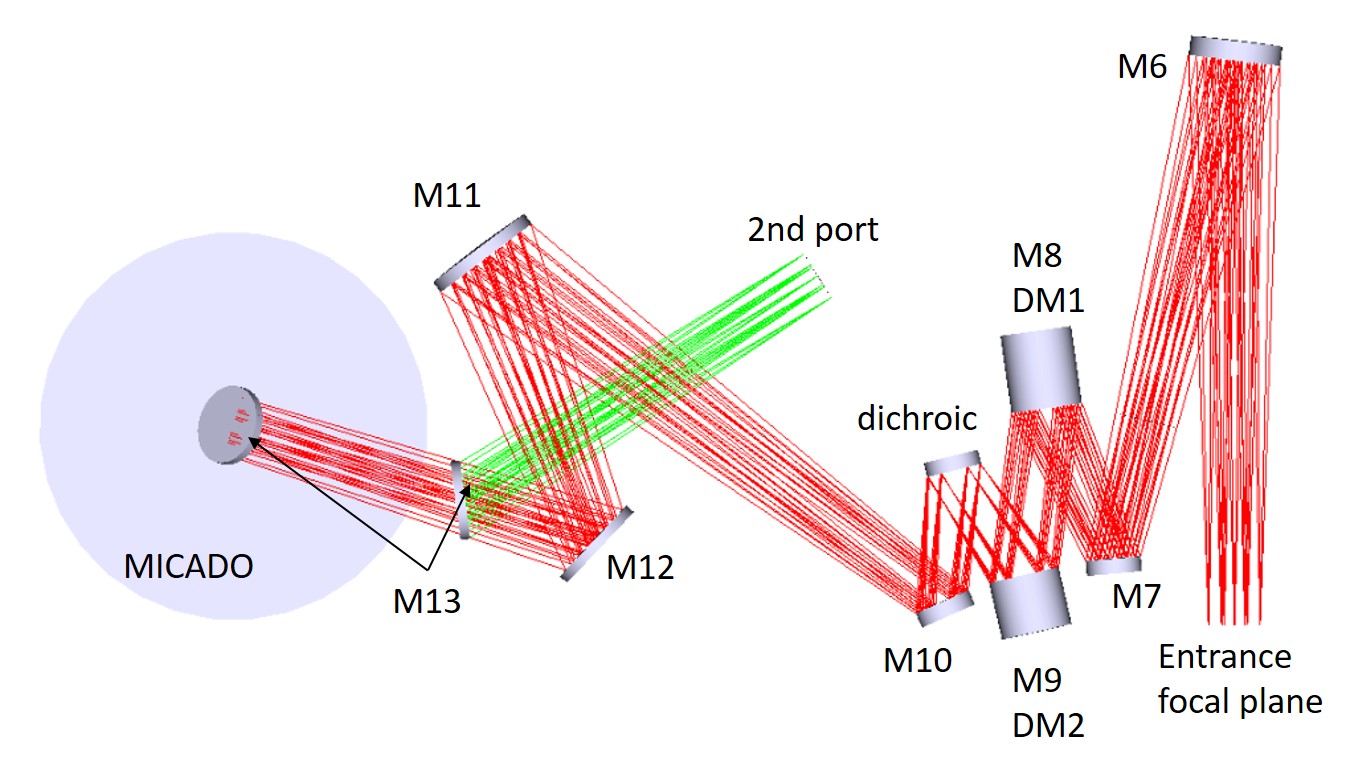}
\caption {Design with reduced number of actuators on DMs.} 
 \label{fig:reduced}
\end{figure}

The MPO consists of 8 mirrors plus the dichroic for MICADO and for the second port (Figure \ref{fig:reduced}).
After the telescope focal plane, the mirrors M6 and M7, both having optical power, create a pupil image where the dichroic is positioned. Before the dichroic the two DMs (M8 and M9) are placed at the layer conjugation ranges of 12.7 km and 5 km respectively. After the dichroic three mirrors with optical power (M10, M11 and M12) produce the exit focal plane preserving the optical interfaces of the telescope focal plane. Flat mirror M13 folds the light downward to create a gravity invariant port for MICADO. The exit port is created by means of a deployable fold mirror (M13).

\begin{table}
\centering
\caption{Smaller DMs design general specifications and  prescriptions.}
	\label{tab:reduces} 
	\begin{tabular}{|l|l|l|l|}
\hline
		\textbf{Surface}   & \textbf{Diame}  &  \textbf{Curv}      & \textbf{Shape}                  \\
                  & [mm]      &                 &                         \\
	           	\hline
        M7        &  1000      &  Concave       &  Ellipsoid    \\
        M8        &  600      &  Convex       &    Sphere                       \\
        M9   &   700      &  Concave         &      Ellipsoid                   \\
        DM@12.7km &            &                &   Ellipsoid                    \\
        M10 &  700      &  Concave       &  Off-axis ellipsoid    \\
        DM@4.7km &            &                &                       \\
        dichroic &   600      &    Flat         &    Sphere                      \\  
        M11  &    700    &    Concave         &      Ellipsoid                    \\
        M12  &    1200    &    Convex         &     Off-axis hyperboloid                     \\
        M13       &  900 x 650     &  Flat        &                         \\
          \hline
\multicolumn{4}{|c|}{\textbf{Optical interfaces to exit port} }                          \\ \hline
\multicolumn{2}{|l|}{Focal ratio} & \multicolumn{2}{l|}{F/17.7} \\ 
\multicolumn{2}{|l|}{Exit pupil distance}   & \multicolumn{2}{l|}{37000 mm (towards telescope)} \\
\multicolumn{2}{|l|}{Focal plane curvature}   & \multicolumn{2}{l|}{9900 mm (convex to telescope)} \\ 
\multicolumn{2}{|l|}{NGS patrol FoV}   & \multicolumn{2}{l|}{160 arcsec diameter} \\ \hline
\end{tabular}
\end{table}

\section{Baseline design}
\label{sect:baseline}

\subsection{Final trade-offs}
\label{sect:finaltradeoffs}

At the end of 2015 ESO discarded, for technical reasons, the option of MICADO being attached to the telescope focus in the risk mitigation path described in Section \ref{sect:reduced}. Therefore the 1:1 relay of the telescope optical interfaces was no more mandatory for the MPO design. MICADO had increased its size 
and it was supposed to be inserted in its final position from the side, through rails. The optimal conjugation altitude of the two MAORY DMs was raised \citep{oberti2017maory} to stay within a range of 14-16 km and 5-8 km respectively.    
A trade-off study was carried out, with the goal of reducing the number of the optical elements in the MPO. Clearly, a smaller number optical elements reduces the possibility to control all the parameters space in the merit function. From the design with 8 mirrors plus dichroic, presented in Section \ref{sect:reduced}, three MPO designs with respectively 7, 6 and 5 mirrors (plus dichroic) were investigated. Comparing the options, eventually it was chosen the 6 mirrors plus dichroic design, which became the baseline design at the beginning of Phase B and is described in detail in the next section. This design has the advantage, respect to the other cases, to deliver the second port at the opposite side of the MAORY bench respect to MICADO position, just inserting a flat folding mirror after the last mirror with optical power (M10 in Figure \ref{fig:baseline}).\\
Since Phase B kick-off in February 2016, two major changes in the interfaces happened: it was decided that MICADO had to be inserted from above and the Nasmyth platform attachment points passed from a 1m x 1m to 3m x 3m grid. From the internal point of view, the MPO optical design was simplified, removing even aspheric terms to the first two mirrors and reducing the asphericity of the DMs, whore curvature radius was imposed to be equal. The different surface asphericity was assumed to be achieved by the actuator themselves. Given the small aspheric deviation, less than 3 microns peak to valley, this solution is well within the range of consolidated technologies.    
  
\subsection{Optical design}
\label{sect:baselineopticaldesign}
\begin{figure}
\begin{center}
\includegraphics[width=\columnwidth]{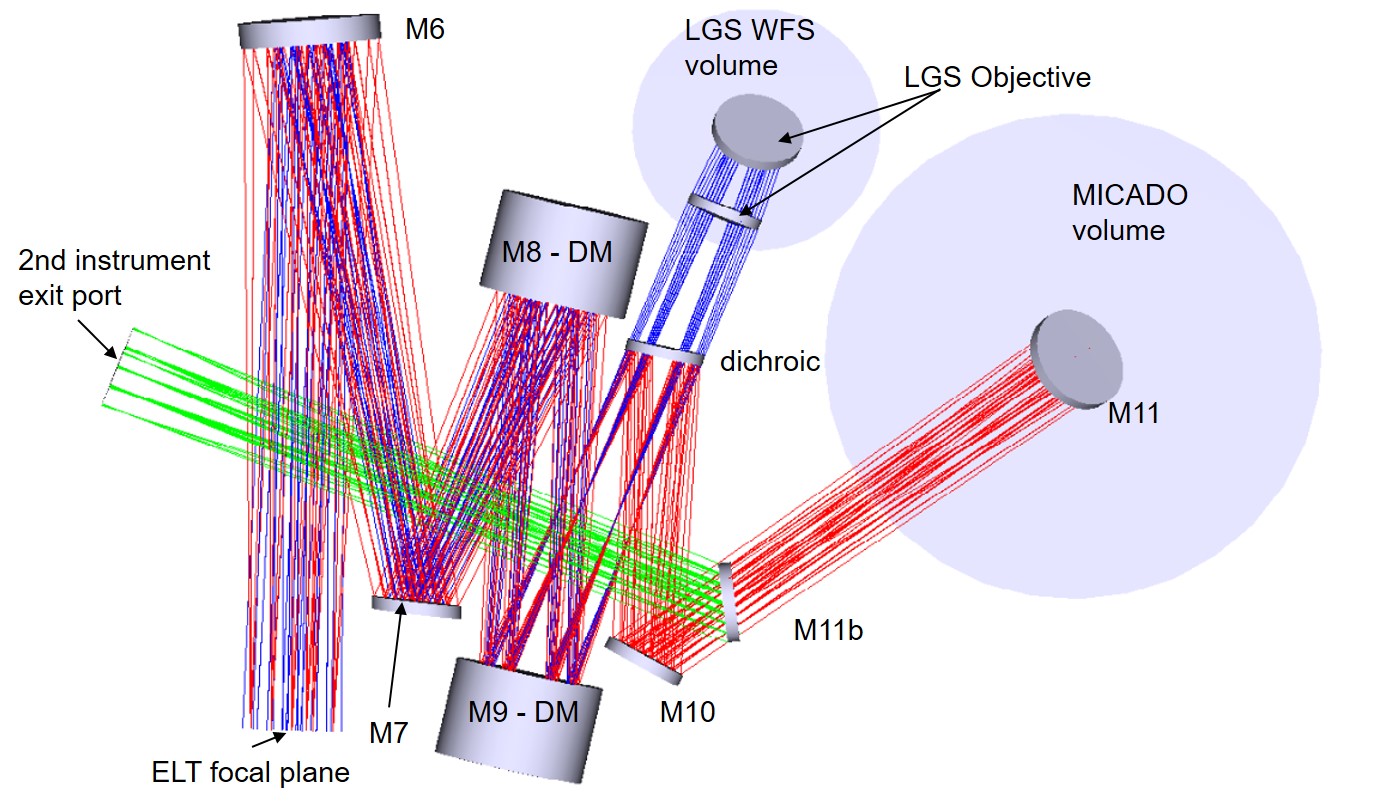}
\end{center}
\caption {Baseline design of the MPO.} 
 \label{fig:baseline}
\end{figure}

The baseline MPO design, shown in Figure \ref{fig:baseline}, consists of 6 mirrors plus the dichroic for MICADO and 7 mirrors plus the dichroic for the second port. It is composed by:
\begin{itemize}
\item a concave off-axis mirror (M6) and convex off-axis mirror (M7), which produce a pupil image of the appropriate size (a concave mirror alone would not be enough, given the available space between the telescope focal plane and the edge of the Nasmyth platform);
\item two concave on-axis DMs (M8 and M9), with the same optical power, optically conjugated to two turbulent high altitude atmospheric layers;
\item the LGS Dichroic, close to the pupil image; this component is assumed to reflect the science light and transmit the LGS light;
\item a convex off-axis mirror (M10), which produces the exit focal plane with the required focal ratio and exit pupil distance;
\item a flat $45^\circ$-tilted mirror (M11), which folds the light to the gravity-invariant port for MICADO. The second instrument port is achieved by inserting a flat folding mirror (M11b) between M10 and M11.
\end{itemize}
The combination of convex mirrors (M7 and M10) and concave mirrors (M7, M8, M9) ensures flat focal surface on the exit port.
The first order parameters and the optical prescription data are listed in Table~\ref{tab:baseline}. The estimated performance and tolerance analysis are described in Section \ref{sect:baseline_perf} and \ref{sect:baseline_tol}.

\begin{table*}
\centering
\caption{Baseline design general specifications and  prescriptions. Concave surfaces are labelled as \textit{cv}, convex surfaces as \textit{cx}.}
	\label{tab:baseline} 
	\begin{tabular}{|l|l|l|l|l|l|l|}
\hline
\textbf{Surface} & \textbf{Diam} & \textbf{Decenter} & \textbf{Tilt}  &\textbf{Curv} & \textbf{Conic} & \textbf{Aspheric} \\
                 &                   &                   &        & \textbf{radius} & \textbf{constant}  &  \textbf{terms} \\
                 & mm                & mm                & degrees &  mm                 &                &       \\ 
\hline
M6               & 1200              & 340               &  6     &13000 (cv)         & 1.9            &             \\ 
M7               & 750               & 800               &  17     &6800 (cx)          & -1             &              \\ 
M8               & 890               &                   &  10     &15000 (cv)         & -5.1           &               \\
DM@15km         &                   &                    &        &                  &                &               \\ 
M9               & 820               &                   &   10    &15000 (cv)         & 2.2            &                \\ 
DM@5km           &                   &                   &        &                  &                &                \\ 
Dichroic         & 600               &                   &    11    &                  &                &                  \\ 
M10              & 675               & 430               &  27     &53400 (cx)         &                & 
\begin{tabular}[c]{@{}l@{}}4th: 6.00e-13\\ 6th: -1.5e-19\\ 8th: 7.5e-26\end{tabular}                              \\ 
M11              & 800 x 600         &                   &    45     &                 &                &                    \\ 
M11              &  700 x 500        &                   &    29    &                  &                &                  \\ 
(for 2nd port)    &                   &                   &        &                  &                &                    \\ 
\hline
\multicolumn{7}{|c|}{\textbf{Optical interfaces to exit port} }                          \\ \hline
\multicolumn{3}{|l|}{Focal ratio} & \multicolumn{4}{l|}{F/17.7} \\ 
\multicolumn{3}{|l|}{Exit pupil distance}   & \multicolumn{4}{l|}{8000 mm (towards telescope)} \\ 
\multicolumn{3}{|l|}{Focal plane curvature}   & \multicolumn{4}{l|}{Flat} \\ 
\multicolumn{3}{|l|}{NGS patrol FoV}   & \multicolumn{4}{l|}{200 arcsec diameter} \\ \hline
\end{tabular}
\end{table*}

\subsection{Performance}
\label{sect:baseline_perf}
\begin{figure}
	\includegraphics[width=\columnwidth]{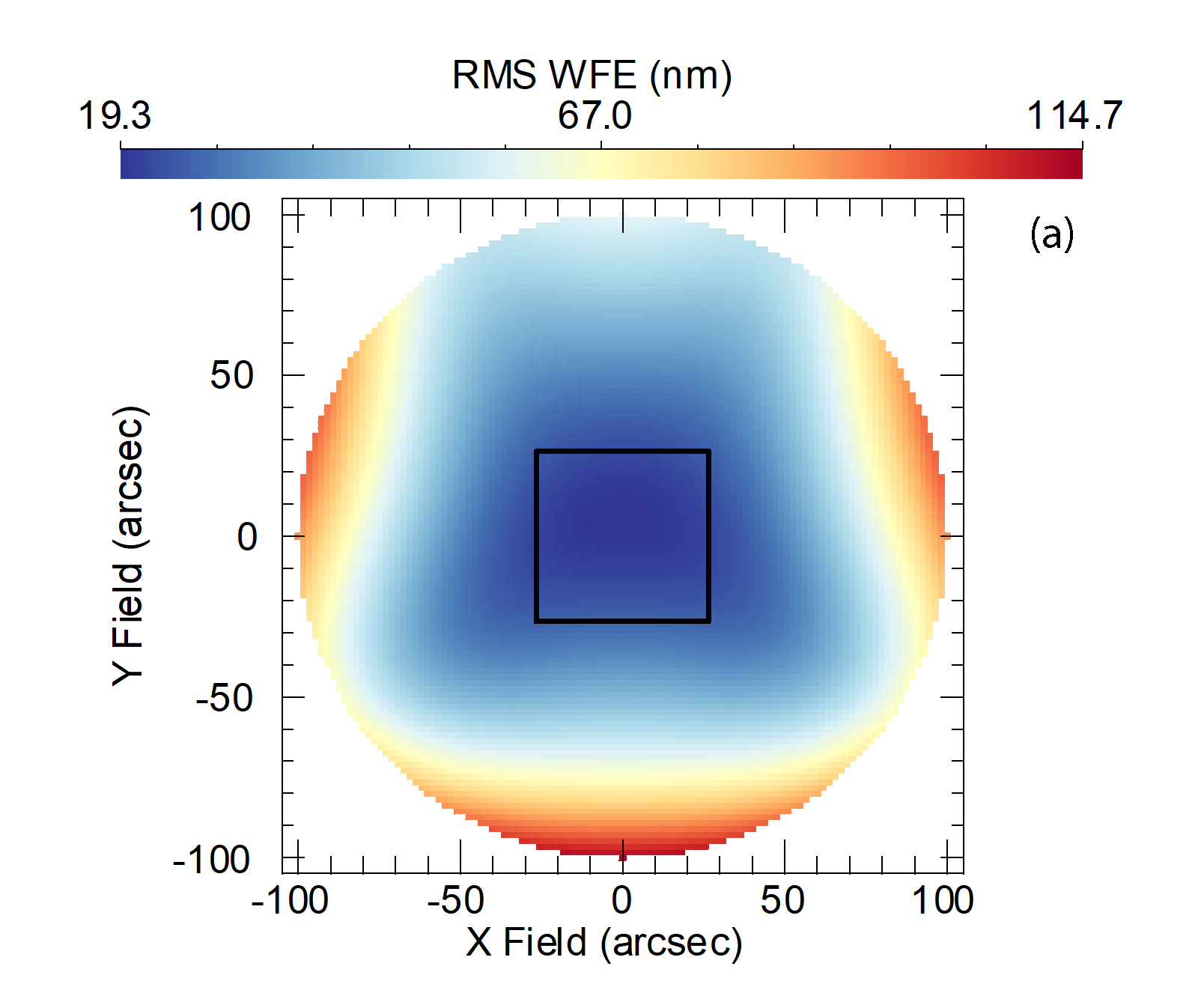}
	\includegraphics[width=\columnwidth]{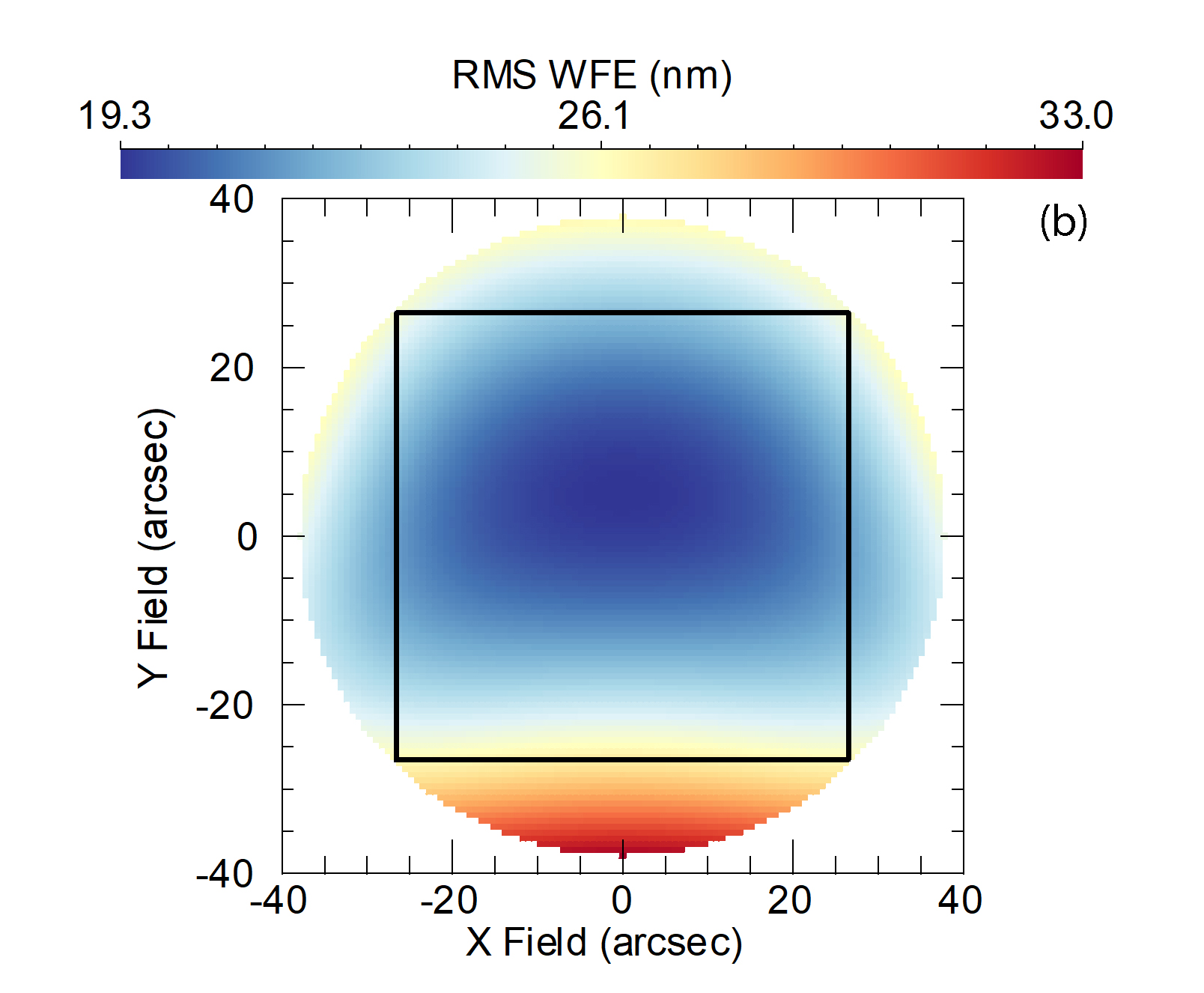}
    \caption {(a): WFE at the exit port of MAORY for the nominal optical design (i.e. without degradation from manufacturing and alignment errors, nominal telescope design included). The NGS patrol FoV is enclosed between the MICADO FoV of 53 x 53 $arcsec^2$ and a 180 arcsec diameter circle. (b): WFE at the exit port of MAORY for the nominal optical design. The MICADO FoV of 53 x 53 $arcsec^2$ is enclosed in the 75 arcsec diameter circle. Wavefront units: nanometers.}
    \label{fig:baseline_wfe}
\end{figure}

\begin{figure}
	\includegraphics[width=\columnwidth]{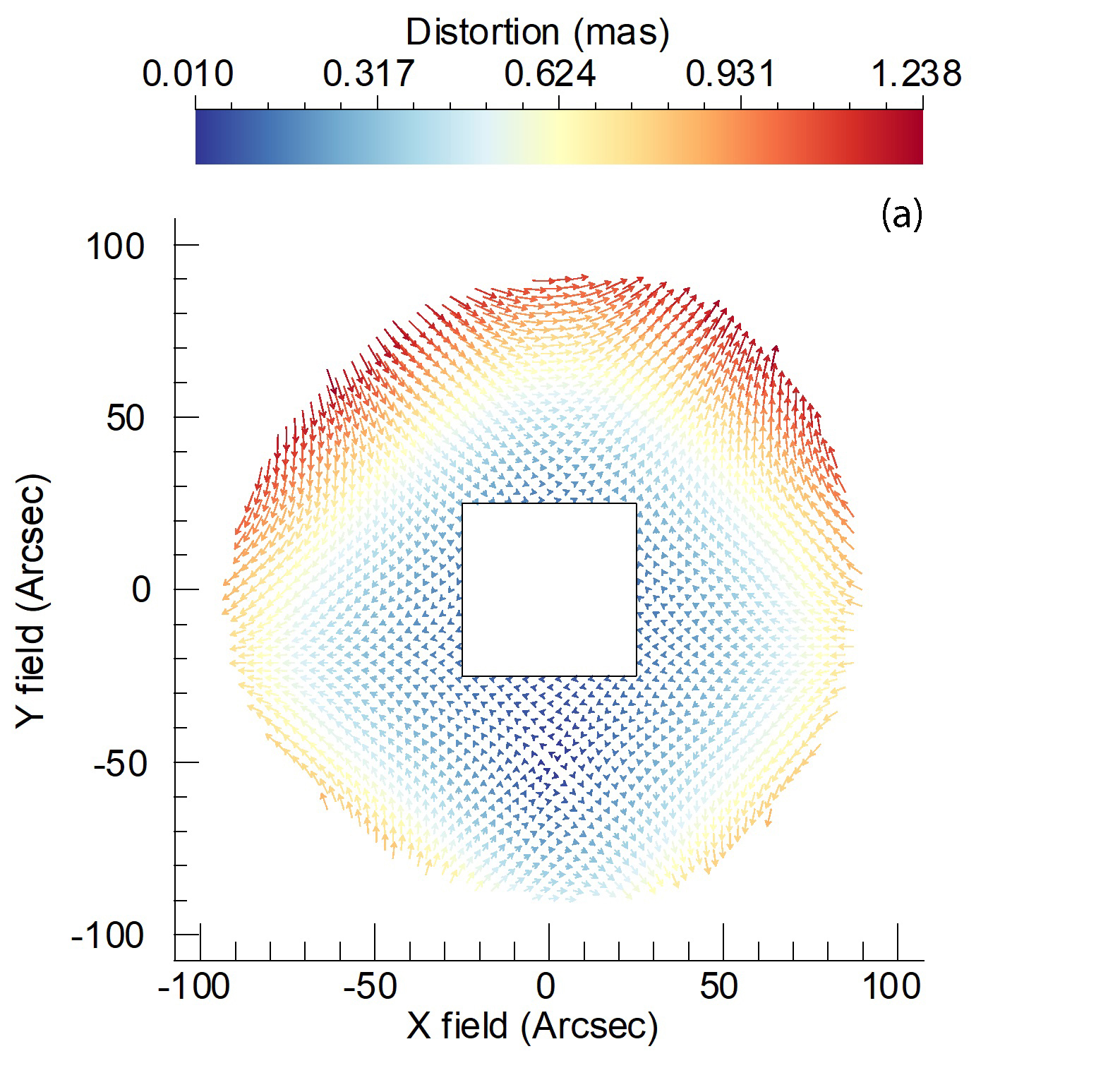}
		\includegraphics[width=\columnwidth]{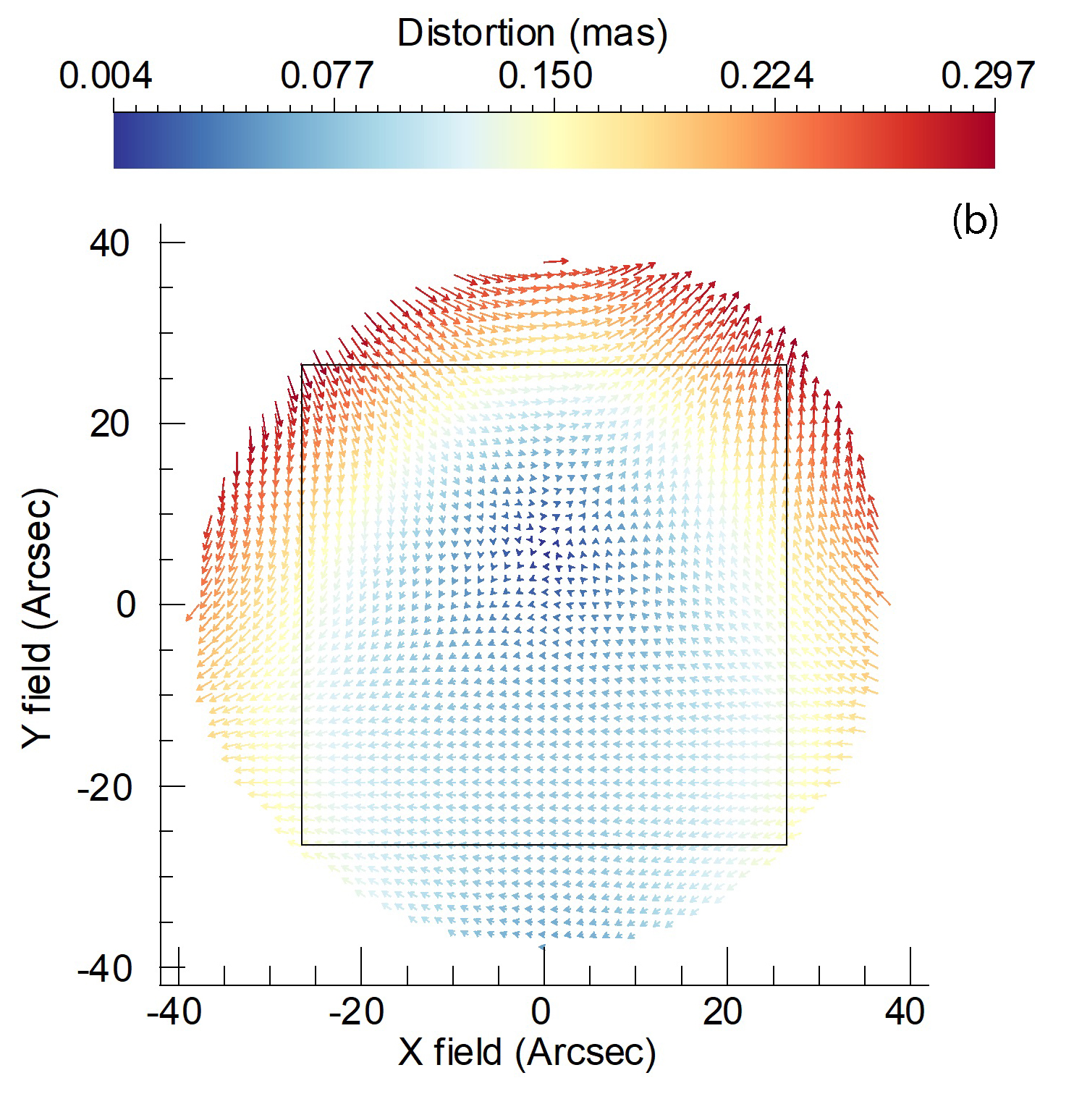}
    \caption {PSFs drift at the MAORY exit port within the maximum integration time for narrow band astrometric observations (without degradation from manufacturing and alignment errors, nominal telescope design included). (a): distortion map in the NGS patrol FoV; (b): distortion map in the MICADO FoV.}
    \label{fig:baseline_distortion}
\end{figure}

\begin{figure}
\begin{center}
\includegraphics[width=\columnwidth]{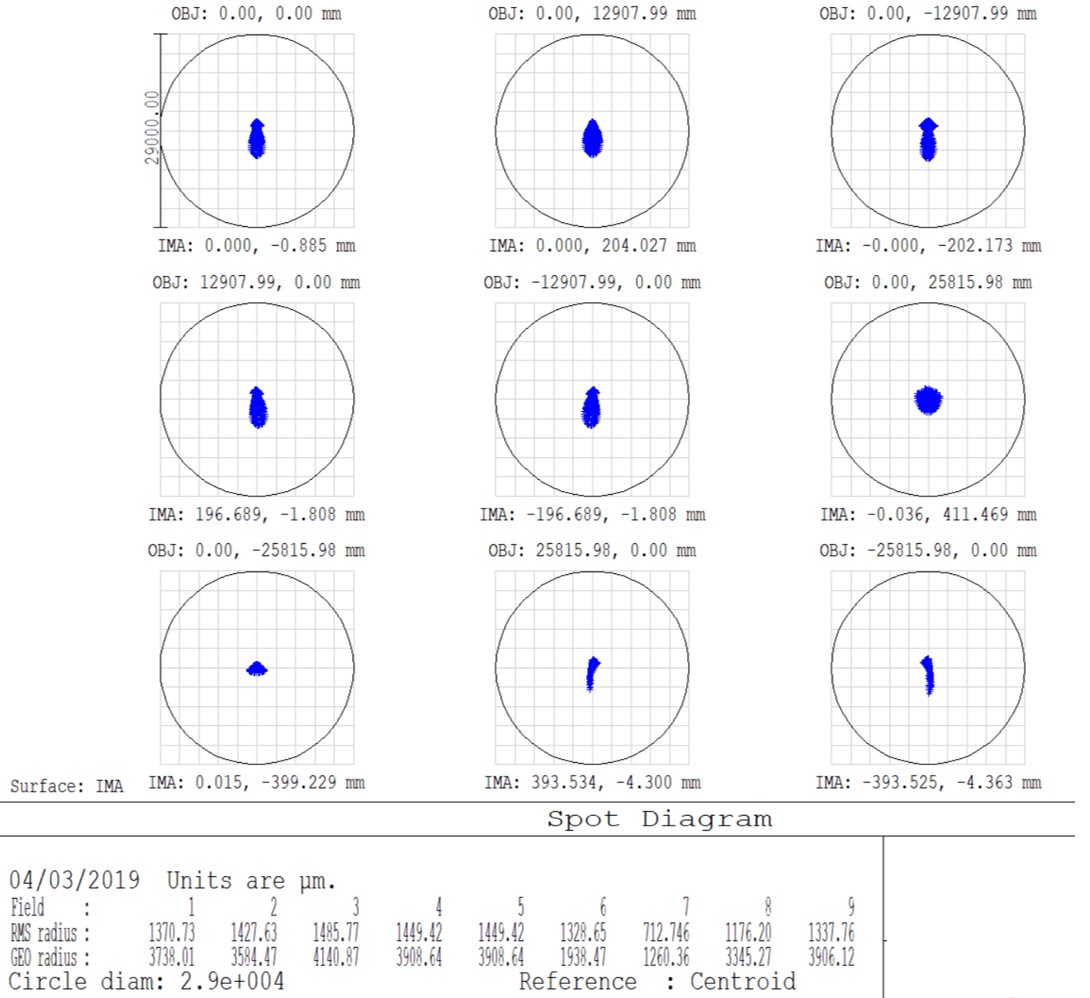}
\end{center}
\caption {Spot diagram for the imagery of a layer at 15 km range on the post-focal DM M8.
The layer is assumed to be parallel to the telescope pupil plane. The circle corresponds to the
actuators pitch.} 
 \label{fig:baseline_DM1_spots}
\end{figure}

\begin{figure}
\begin{center}
\includegraphics[width=\columnwidth]{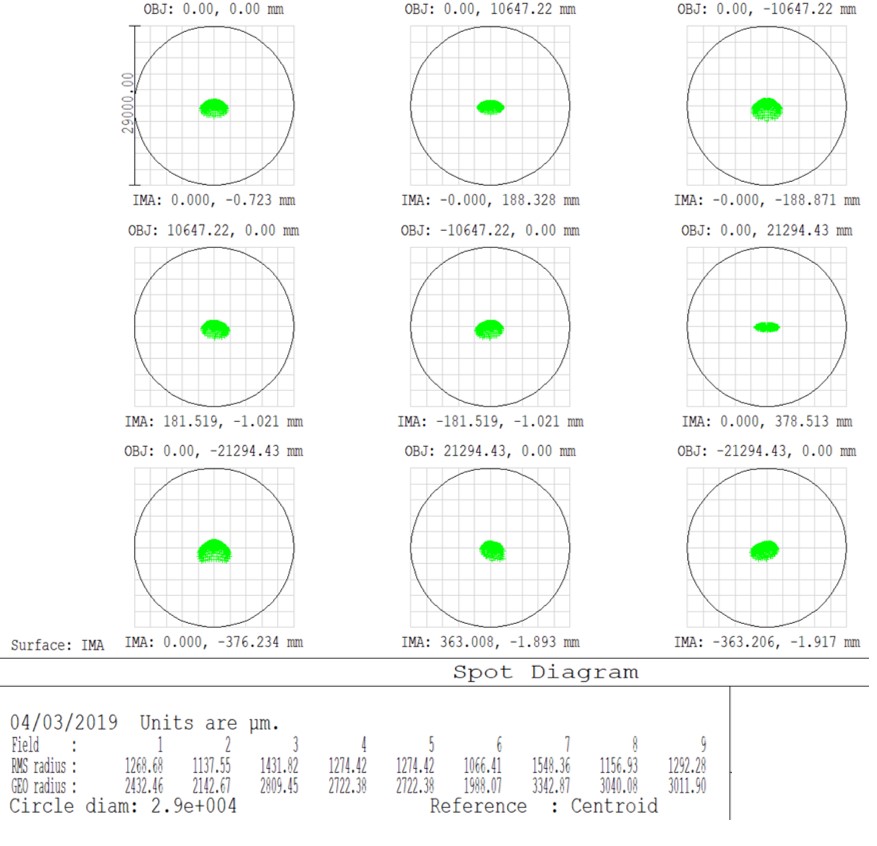}
\end{center}
\caption {Spot diagram for the imagery of a layer at 5 km range on the post-focal DM M9.
The layer is assumed to be parallel to the telescope pupil plane. The circle corresponds to the
actuators pitch.} 
 \label{fig:baseline_DM2_spots}
\end{figure}

The RMS WFE at the exit focal plane of the MPO, for the baseline design, is shown in Figure \ref{fig:baseline_wfe}. In particular, the WFE is below 30 nm RMS over the MICADO FoV. \\
Regarding the geometric distortion performance, explained in Section \ref{sect:dist} and shown for the baseline design in Figure \ref{fig:baseline_distortion}, it is well below the requirements.\\
The optical quality of the post-focal relay exit pupil, i.e. as it appears from an instrument point of view, has been tested by a paraxial lens placed after the relay. The RMS radius of the blur on the pupil image, formed by all the rays over the MICADO FoV is 1/2000 of pupil diameter. The optical quality of the layer images on the DMs, in terms of RMS spot radius, is shown in Figure \ref{fig:baseline_DM1_spots} and \ref{fig:baseline_DM2_spots} for the DM1 and DM2 respectively. In these spot diagrams the layer are assumed to be parallel to the telescope pupil plane. The spots are in the order of 1/10 DMs pitch size.

\subsection{Tolerances}
\label{sect:baseline_tol}

\begin{table*}
\centering
\caption{MPO manufacturing tolerances. Spherical aberration for flat surfaces is included in the budget of irregularities.}
	\label{tab:manuftol1} 
\begin{tabular}{|l|l|l|l|l|l|}
\hline
\multicolumn{6}{|c|}{\textbf{Manufacturing tolerances}}   \\ \hline
\multicolumn{3}{|l|}{Curvature radius}   & \multicolumn{3}{l|}{$\pm0.1 \%$}   \\ 
\multicolumn{3}{|l|}{\begin{tabular}[c]{@{}l@{}}Sag of flat surfaces (residual curvature))\end{tabular}} & \multicolumn{3}{l|}{$\pm 316$ nm}   \\ \hline
\multicolumn{6}{|c|}{\textbf{Spherical aberration (RMS)}}  \\ \hline
\multicolumn{3}{|l|}{All mirrors with optical power}      & \multicolumn{3}{l|}{5 nm}   \\ \hline
\multicolumn{6}{|c|}{\textbf{Surface irregularity (RMS)}}   \\ \hline
Mirror       & Astigmatism      &       Zernike range     &  Zernike range          &   Zernike range      &  Zernike range \\
             & Coma + Trefoil  &    $12 \leq z \leq 28$      &  $29 \leq z \leq 45$      & $46 \leq z \leq 120$      & $121 \leq z \leq 230$ \\
\hline
M6           &    24 nm       &  5 nm                     & 5 nm                    &   4 nm               &                 4 nm \\
M7           &    18 nm       &  5 nm                     & 3 nm                    &   4 nm               &                 4 nm \\
M8           &    15 nm       &  5 nm                     & 5 nm                    &   4 nm               &                 4 nm \\
M9           &   14 nm       &  5 nm                     & 3 nm                    &   4 nm               &                 4 nm \\
M10           &    20 nm       &  5 nm                     & 3 nm                    &   4 nm               &                 4 nm \\
M11           &    23 nm       &  5 nm                     & 5 nm                    &   4 nm               &                 4 nm \\
Dichroic     & 11 nm       &  5 nm                     & 5 nm                    &   4 nm               &                 4 nm \\
\hline
\end{tabular}
\end{table*}

\begin{table}
\centering
\caption{Tolerances of Main Path Optics degrees of freedom. Required accuracy during system mechanical assembly and required stability after system alignment achieved with compensators, described in Section \ref{sect:baseline_align}.}
	\label{tab:manuftol2} 
\begin{tabular}{|l|l|}
\hline
\multicolumn{2}{|c|}{\textbf{Tolerances of stability}}   \\ 
\hline
X-Y Tilts (M7 - M9 - M10 - Dichroic) &  $\pm 30$ $\mu$rad \\
X-Y Tilts (M6)&	 $\pm 87$ $\mu$rad  \\
X-Y Tilts (M8 - M11)& $\pm 18$ $\mu$rad  \\
Z Tilt (All optics rotation around optical axis)  & $\pm 30$ $\mu$rad  \\
Axial position (M8 repeatability)  & $\pm 0.1$ mm 	\\
Axial position (All optics repeatability except M8)	& $\pm 0.2$ mm \\
Axial position (compensated by re-focus) & $\pm 1$ mm \\
X-Y Decenters (All optics) &	 $\pm 0.1$ mm \\
\hline
\multicolumn{2}{|c|}{\textbf{Tolerances of assembly}}   \\ 
\hline
X-Y Tilts &	$\pm 1$ $\mu$rad  \\
Axial position & $\pm 1$ mm \\	
X-Y Decenters &	  $\pm 0.25$ mm \\
Z Tilt (rotation around optical axis)&	 $\pm 1$ $\mu$rad \\
\hline
\end{tabular}
\end{table}
 
The method used to estimate tolerances takes care of compensation of errors during assembly or alignment procedure and uses a root-sum-square (RSS) approach to combine independent error contributions~\citep{patti2018maory}. The sensitivity analysis on system performance considers each tolerance individually and, once a merit function is defined, the errors are combined by RSS to find the net effect of all the tolerances on the system. This method assumes that errors introduced by different tolerances are statistically uncorrelated and allows to identify parameters which are highly sensitive to certain errors, such as surface curvature radii or decenters. The sensitivities related only to available degree-of-freedom (DOF) can be used during the alignment phase of the instrument since the worst offenders to system performance are also the best set of compensators for required adjustments. 
There are two requirements that limit the tolerable errors of opto-mechanical parameters: the FoV-averaged RMS WFE and the geometric distortion.    
Considering the RMS WFE, the goal was to maintain the maximum WFE on all the MICADO FoV below the diffraction limit @1 $\mu$m wavelength (70 nm RMS) while regarding the geometric distortion the goal was to remain below the requirements described in Section~\ref{sect:dist}. The tolerance analysis has been broken down into 2 blocks that consider different error sources of the optical elements.
Block 1 is the manufacturing tolerance and can be split into 2 sub blocks: curvature radii and low order surface irregularities ($4 < z \leq 11$) and high order surface irregularities ($12 \leq z \leq 230$), which are modelled by standard Zernike coefficients \citep{noll1976zernike}. The results are summarized in Table \ref{tab:manuftol1}. Precision optics are required to maintain the WFE degradation below the requirements for the fist sub block. Moreover a partial compensation can be done re-optimizing the optical design once the real parameters are measured after the manufacturing. High order surface irregularities mainly affect the astrometric performance, since the MPO does not rotate as the sky. In particular M11, the closest mirror to the exit port, is the most critical element. \\
Block 2 considers the mechanical tolerances and sets the precision and stability of the mounts and alignment stages. Regarding the assembly, the laser tracker accuracy (Section \ref{sect:baseline_align}) is sufficient for our requirements and the high precision commercial stages can fulfill the requirements regarding the repeatability. Results are summarized in Table \ref{tab:manuftol2}.\\

\subsection{Alignment concept}
\label{sect:baseline_align}

\begin{figure}
\begin{center}
\includegraphics[width=\columnwidth]{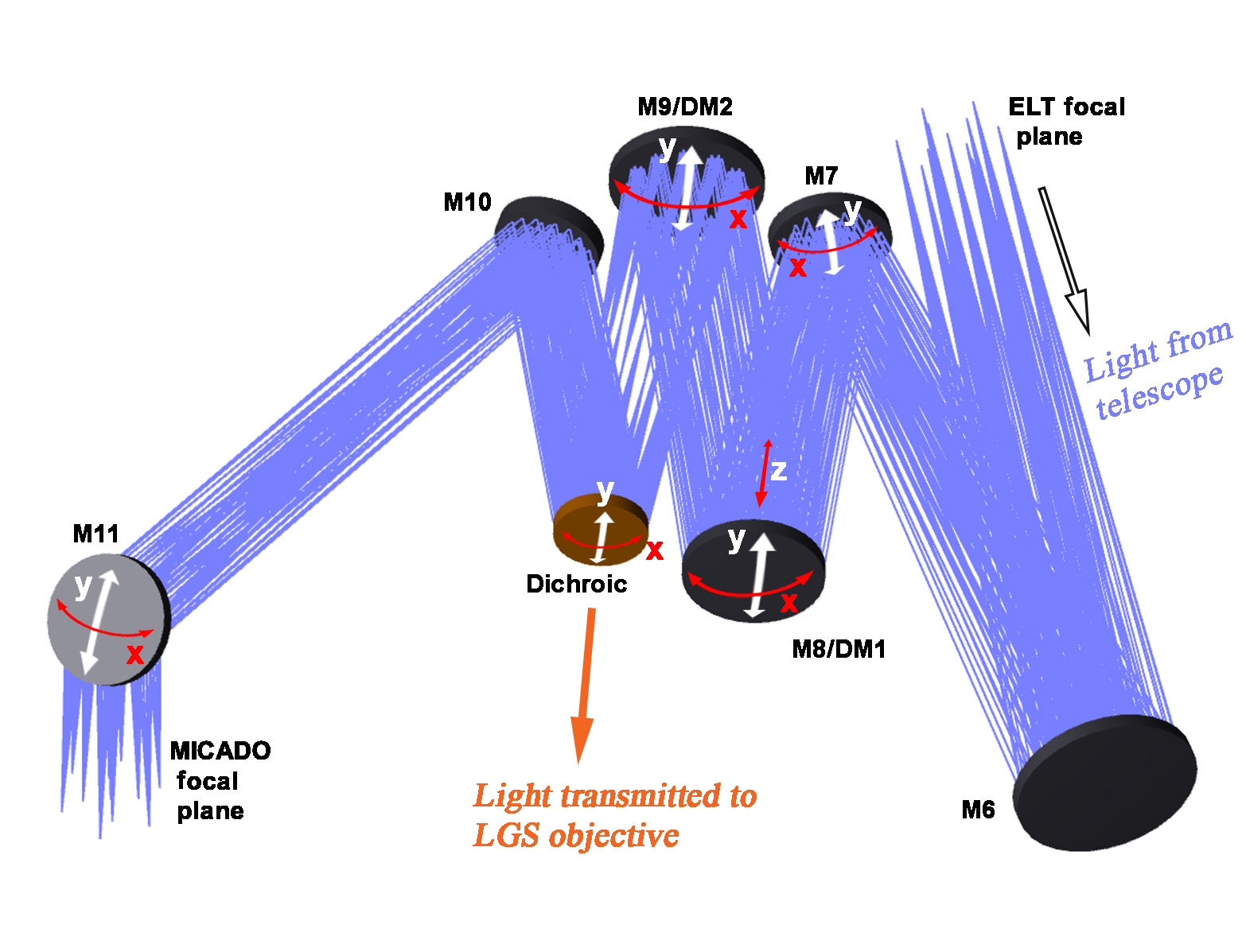}
\end{center}
\caption 
{MAORY MPO and its DOF compensators. Tilts on local x and y axes are necessary compensators to achieve diffraction limited optical quality during the alignment (plus one mirror axial distance Z). M11 Tilts are additional compensators to achieve the requirement on the geometrical distortion. } \label{fig:align}
\end{figure} 
The MPO alignment plan is foreseen to be carried on in two steps: 
1. Optics, entrance focal plane (where to put optical fibers) and exit focal plane (where to put cameras and WFSs) are positioned with laser tracker accuracy; 
2. The final alignment.
The sensitivity analysis identified the worst offender DOFs that are used as active compensators (Figure \ref{fig:align}). They should work in combination with the observables at the focal plane (WFE and images position) that provide the necessary information to change the optical path according to estimated mis-alignments. A Monte Carlo simulation has been carried out for the entire procedure. For each case it is assumed to use some sources at the entrance focal plane and measure both positions and WFE at the exit focal plane. Zernike coefficients and distortion values are target values of the merit function of the nominal design. The optimization is done using the damped least squared algorithm and the compensators as variables. The perturbed DOFs of the compensators will be the applied motions (in a reverse way) to align the system. This method, called Reverse Optimization (RO), is described in detail in \citet{patti2019prep}. \\
The results of the RO method show that the mean residual WFE in the MICADO FoV is few nm bigger than the nominal design. Regarding the geometrical distortion, the error is about 1 milli-arcsecond at the edge of the MICADO FoV. Some possible errors in the measurement of the spots centroids and of the WF, which are expected to be present in the real system during both the alignment phase in the lab and the commissioning at the telescope, have been added to this procedure. 

\subsection{Back-up solution} 
\label{sect:backup}

The dichroic is assumed to be coated with a low bandwidth filter that transmits below 600 nm wavelength and reflects longer wavelengths up to 2400 nm. This choice ensures the achromaticity of the MPO and a minor risk on the substrate material, which must transmit the almost monochromatic LGS light. Anyway, coating efficiency is better established for the reverse behaviour, i.e. IR-transmitting. As one of the back-up solutions, a MPO design with the dichroic transmitting the science light has been investigated and produced. The residual WFE and geometric distortion for MICADO were still within the requirements and the lateral chromatism due to the tilted refractive element was acceptable. The drawbacks were the increased thermal background due to the dichroic size that required, as substrate, infrared-grade fused silica material and the need of two deployable mirrors to produce the second exit port. After further interactions with Optics manufacturers, it has been decided to continue with the baseline design.\\

\section{Conclusions}

We presented the historical evolution of the MPO design of MAORY until Spring 2018. Further little modifications of the design are beyond the scope of this paper. For the convergence to the baseline design, many possible paths, most of them not described in this paper, has been taken and eventually discarded for not fulfilling all the requirements or because of other factors as cost, complexity, etc. Part of the difficulties of this design work come from the relatively small envelope assigned to MAORY. In fact on the Nasmyth platform the four first light instruments must live together, differently from the 8-m class telescopes where one instrument at the time is installed. The optics size scales linearly with the telescope diameter, but not the assigned space, thus the optics must be necessarily more complicated, i.e fast and off-axis conic surfaces. 
Nevertheless the MPO baseline design appears solid from both the manufacturing and the alignment point of view.




\bibliographystyle{mnras}
\bibliography{Lombini_MNRAS_2019_revised} 

\bsp	
\label{lastpage}
\end{document}